\documentclass[10pt]{article}
\usepackage{amsmath, amssymb, amsthm, amscd}
\usepackage{amsrefs}
\usepackage{graphics}

\newcommand{\binomial}[2]  {{\left(\begin{array}{c}#1\\#2\end{array}\right)}}

\newtheorem{theorem}      {Theorem}
\newtheorem{restatedtheorem}{Theorem}

\newtheorem*{theorem*}    {Theorem}

\newtheorem{lemma}        {Lemma}

\newtheorem{proposition}  {Proposition}
\newtheorem{corollary}    {Corollary}

\newtheorem{algorithm}    {Algorithm}

\theoremstyle{definition}
\newtheorem{definition}   {Definition}

\newtheorem{problem}      {Problem}

\newcommand{\CompClass}[1]{\ensuremath{\mathbf {#1}}}
\newcommand{\NPO}{\CompClass{NPO}}

\newcommand{\APX}{\CompClass{APX}}

\newcommand{\NP}{\CompClass{NP}}
\newcommand{\BPP}{\CompClass{BPP}}
\newcommand{\Po}{\CompClass{P}}
\newcommand{\slashp}{\CompClass{\# P}}

\newcommand{\bez}{\mathrm{B\acute ez}}
\newcommand{\vol}{\mathrm{Vol}}
\newcommand{\conv}{\mathrm{Conv}}

\newcommand{\boldalpha}{\boldsymbol{\alpha}}
\newcommand{\partition}[1]{\boldsymbol #1}

\newcommand{\bydef}{\ensuremath{\stackrel{\mbox{\scriptsize def}}{=}}}

\begin{document}
\author{Gregorio Malajovich\footnote{
{\bf Postal address:} 
Departamento de Matem\'atica Aplicada,
Universidade Federal do Rio de Janeiro.
Caixa Postal 68530, CEP 21945-970, Rio de Janeiro,
RJ, Brasil. {\bf email:}{\tt gregorio@ufrj.br}.
{\bf url:} {\tt www.labma.ufrj.br//\~{ }gregorio}.
{\bf Acknowledgements:} G.M. is partially supported
by CNPq (Brasil). This 
work was done while visiting
Syddansk Universitet at Odense, thanks to the
generous support of the
Villum Kann Rasmussen Fond.} \\
\and Klaus Meer
\footnote{{\bf Postal address:}
Department of Mathematics and Computer Science.
University of Southern Denmark,
Campusvej 55,
DK-5230, Odense M.
Denmark.
{\bf email:}{\tt meer@imada.sdu.dk}.
{\bf url:} {\tt www.imada.sdu.dk//\~{ }meer}.
{\bf Acknowledgements:}
Partially supported 
by the EU Network of Excellence PASCAL 
Pattern Analysis, Statistical Modelling and Computational Learning
and by the Danish Natural Science Research Council SNF.}}

\title{Computing Multi-Homogeneous B\'ezout Numbers is Hard}

\date{29 april 2004}
\maketitle

\begin{abstract}
The multi-homogeneous B\'ezout number is a bound for
the number of solutions of a system of multi-homogeneous
polynomial equations, in a suitable product of projective
spaces.

  Given an arbitrary, not necessarily multi-homogeneous 
system, one can ask for the optimal multi-homogenization
that would minimize the  B\'ezout number.

  In this paper, it is proved that the problem of computing,
or even estimating the optimal multi-homogeneous B\'ezout 
number is actually $\NP$-hard. 

  In terms of approximation theory for combinatorial 
optimization, the problem of computing the best
multi-homogeneous structure does not belong to $\APX$,
unless $\Po = \NP$.

  Moreover, polynomial time algorithms for estimating
the minimal multi-homo\-ge\-neous B\'ezout number up to a
fixed factor cannot exist even in a randomized setting,
unless $\BPP \supseteq \NP$.
\end{abstract}

\section{Introduction}

The multi-homogeneous B\'ezout number is a bound for
the number of solutions of a system of multi-homogeneous
polynomial equations.

Estimating the number of isolated solutions of a polynomial
system is useful for the design and analysis of
homotopy algorithms~\cite{LI}. Applications include 
problems in engineering like the design of certain mechanisms
~\cites{MORGAN, WMS} or others, such as computational
geometry.

An application of 
multi-homogeneous B\'ezout bounds
outside the realm
of algebraic equation solving is discussed in~\cite{DMS}, 
where the number of roots
is used to bound geometrical quantities such as
volume and curvature.

There is an important connection between root-counting and
$\NP$-completeness theory. Indeed, it is easy to
reduce an $\NP$-complete or $\NP$-hard 
problem such as SAT, the Traveling
Salesman problem, Integer Programming (and thus all 
other
$\NP$ problems as well) to the question whether certain 
polynomial
systems have a common zero. 

\bigskip
\par

The best-known example giving an estimate for the number of roots
of a polynomial equation is the 
Fundamental Theorem of Algebra. It was generalized 
to multivariate polynomial systems  at the 
end of the 18th century by Etienne B\'ezout. The B\'ezout number
bounds the number of (isolated) complex solutions of a polynomial
$f : \mathbb C^n \mapsto \mathbb C^n$ from above by the product of the degrees 
of the involved polynomials. However, in many cases this estimate is 
far from optimal. A well known example is given by the eigenvalue
problem: Given a $n \times n$
matrix $M$, find the eigenpairs $(\lambda, u) \in
\mathbb C \times \mathbb C^n$ such that 
$M u - \lambda u = 0$. If we
equate $u_n$ to $1$,
the classical B\'ezout number becomes $2^{n-1},$ though of course only 
$n$ solutions exist.

The multi-homogeneous B\'ezout number provides a sharper bound on
the number of isolated solutions of a system of equations,
in a suitable product of projective spaces.
The multi-homogeneous B\'ezout bound depends on
the choice of a {\em multi-homogeneous structure},
that is of a partition of the variables $(\lambda, u)$
into several groups.

In the eigenvalue example, 
the eigenvector $u$ is defined up to a multiplicative
constant, so it makes sense to define it as an element
of $\mathbb P^{n-1}$. With respect to the eigenvector
$\lambda$, we need to introduce a homogenizing variable.
We therefore rewrite the equation as:
$\lambda_0 M u - \lambda_1 u = 0$, and $\lambda = 
\lambda_1 / \lambda_0$. Now the pair $(\lambda_0 : 
\lambda_1)$ is an element of $\mathbb P^1$. The
multi-homogeneous B\'ezout number for this system
is precisely $n$.

%
%
%
%
%

\medskip

Better bounds on the root number
are known, such as
Kushnirenko's~\cite{KUSHNIRENKO} or Bernstein's
~\cite{BERNSTEIN}.
However, interest in computing the multi-homogeneous B\'ezout
number stems from the fact that hardness results are
known for those sharper bounds
(see section~\ref{sec:volume}
for details).

Another reason of interest is that in many cases, a
natural multi-homogeneous structure is known or may
be found with some additional human work.

\medskip

In this paper, we consider the following problem.
{\em 
Let $n \in \mathbb N$ and a finite $A \subset \mathbb N^n$
be given as input. Find the minimal multi-homogeneous
B\'ezout number, among all choices of a
multi-homogeneous structure for a polynomial system
with support $A$:}
\begin{equation}
\label{eq:*}
\left\{
\begin{array}{lcl}
f_1(z) &=& \sum_{\mathbf \boldalpha \in A} f_{1 \mathbf \boldalpha} 
z_1^{\alpha_1} z_2^{\alpha_2} \cdots z_n^{\alpha_n} \\
& \vdots & \\
f_n(z) &=& \sum_{\mathbf \boldalpha \in A} f_{n \mathbf \boldalpha} 
z_1^{\alpha_1} z_2^{\alpha_2} \cdots z_n^{\alpha_n} \ .\\
\end{array}
\right.
\end{equation}

where the $f_{i \mathbf \alpha}$ are non-zero complex coefficients.

Geometrically, this minimal B\'ezout number is an
upper bound for the number of isolated roots of the
system~(\ref{eq:*}) in $\mathbb C^n$.
\medskip

The main result in this paper (restated formally  in section~\ref{sec:bezout}
below) is:

\begin{theorem}\label{th:A}
There cannot possibly exist a polynomial
time algorithm to approximate the minimal multi-homogeneous
B\'ezout number for (\ref{eq:*}) up to any fixed factor, 
unless $\Po = \NP$.
\end{theorem}

This means that computing or even approximating the
minimal B\'ezout number up to a fixed factor is 
\NP-hard. In terms of the hierarchy of approximation
classes (see \cite{AUSIELLO} and section~\ref{sec:approx}),
the minimal multi-homogeneous B\'ezout number does not
belong to the class $\APX$ unless $\Po = \NP$.
\medskip

Motivated by what is known on volume approximation
(see section~\ref{sec:volume}), one could 
ask whether allowing for randomized algorithms would
be of any improvement.

\begin{theorem}\label{th:B}
There cannot possibly exist a randomized
polynomial time algorithm to approximate the 
minimal multi-homogeneous
B\'ezout number for (\ref{eq:*}) up to any fixed factor,
with probability of failure $\epsilon < 1/4$,
unless $\BPP \supseteq \NP$.
\end{theorem}

While the conjecture $\BPP \not \supseteq \NP$ is less
widely known outside the computer science community
than the conjecture $\Po \neq \NP$, its
failure would imply the existence of probabilistic 
polynomial time algorithms for solving problems such
as the factorization of large integers or the 
discrete logarithm. Most widespread cryptographic
schemes are based on the assumption that those two
problems are hard.

\section{Background and Statement of Main Results}
\label{sec:background}
\subsection{B\'ezout numbers}
\label{sec:bezout}

In the definition of (\ref{eq:*}), we assumed
for simplicity that each equation had the same
{\em support} $A$. In general, a system $f(z)$ of
$n$ polynomial equations with support $(A_1, \dots, A_n)$
is a system of the form:

\begin{equation}
\label{eq:**}
\left\{
\begin{array}{lcl}
f_1(z) &=& \sum_{\boldalpha \in A_1} f_{1 \mathbf \boldalpha} 
z_1^{\alpha_1} z_2^{\alpha_2} \cdots z_n^{\alpha_n} \\
& \vdots & \\
f_n(z) &=& \sum_{\boldalpha \in A_n} f_{n \mathbf \boldalpha} 
z_1^{\alpha_1} z_2^{\alpha_2} \cdots z_n^{\alpha_n} \ ,\\
\end{array}
\right.
\end{equation}
where the coefficients $f_{i \mathbf \boldalpha}$ are non-zero
complex numbers.

A {\em multi-homogeneous structure} is given by a 
partition of $\{1, \dots , n\}$ into (say) $k$ sets
$I_1, \dots, I_k$. Then for each set $I_j$, we consider
the group of variables $Z_j = \{ z_i : i \in I_j \}$.

The degree of $f_i$ in the group of variables $Z_j$ is
\[
d_{ij} \bydef \max_{\boldalpha \in A_i} \ \sum_{l \in I_j}
\ \alpha_l
\]

When for some $j$, for all $i$, the maximum $d_{ij}$ is
attained for all $\boldalpha \in A_i$, we say that
(\ref{eq:**}) is homogeneous in the variables $Z_j$.
The dimension of the projective space associated to
$Z_j$ is:
\[
a_j \bydef \left\{
\begin{array}{ll}
\# I_j - 1& \text{if (\ref{eq:**}) is homogeneous in $Z_j$, and}
\\
\# I_j & \text{otherwise.}
\end{array}
\right.
\]

We assume that $n = \sum_{j=1}^k a_j$. Otherwise,
we would have an undetermined ($n < \sum_{j=1}^k a_j$)
or overdetermined ($n > \sum_{j=1}^k a_j$) 
polynomial system, and multi-homogeneous B\'ezout
numbers would have no meaning.

The multi-homogeneous B\'ezout number 
$\bez(A_1, \dots, A_n; I_1, \dots, I_k)$
is the coefficient of 
$\prod_{j=1}^k \zeta_j^{a_j}$ in the formal
expression $\prod_{i=1}^n \sum_{j=1}^k d_{ij} \zeta_j$
(see~\cites{MORGAN-SOMMESE,LI,SHAFAREVICH}). It bounds 
the maximal number of
isolated roots of (\ref{eq:**}) in
$\mathbb P^{a_1} \times \cdots \times \mathbb P^{a_k}$.
Therefore it also bounds the number of {\em finite} roots
of (\ref{eq:**}), i.e. the roots in $\mathbb C^n$.

\medskip

In the particular case where $A = A_1 = \cdots = A_n$
there is a simpler expression for the multi-homogeneous
B\'ezout number $\bez(A;I_1, \dots, I_k) \bydef
\bez(A_1, \dots, A_n;$ $I_1, \dots, I_k)$, namely:
\begin{equation}\label{eq:bezfromI}
\bez(A;I_1, \dots, I_k) =
\binomial {n}{a_1\ a_2\ \cdots\ a_k} 
\
\prod_{j=1}^k d_j^{a_j} \ ,
\end{equation}

where $d_j = d_{ij}$ (equal for each $i$) and the 
multinomial coefficient 
\[
\binomial {n}{a_1\ a_2\ \cdots\ a_k} 
\bydef
\frac{n!}{a_1!\ a_2!\ \cdots\ a_k!}
\]
is the coefficient of $\prod_{j=1}^k \zeta_j^{a_k}$
in $(\zeta_1 + \cdots + \zeta_k)^{n}$ (recall that
$n = \sum_{j=1}^k a_j$). 

\medskip
\par

Heuristics for computing a suitable multi-homogeneous
structure $(I_1, \dots, I_k)$ given $A_1, \dots, A_n$
are discussed in \cites{LI-BAI, LI-LIN-BAI}. Surprisingly
enough, there seems to be no theoretical results available
on the complexity of computing the minimal B\'ezout number.
It was conjectured in \cite[p.78]{LI-BAI} that 
computing the minimal multi-homogeneous B\'ezout
number is $\NP$-hard. 

Even, no polynomial time algorithm for computing the multi-
homogeneous B\'ezout number {\em given a multi-homogeneous
structure} seems to be known (see~\cite[p.240]{LI-LIN-BAI}).

This is why in this paper, we restrict ourselves to
the case $A=A_1 = \cdots = A_n$. This is a particular
subset of the general case, and any hardness result for
this particular subset implies the same hardness 
result in the general case.

More formally, we adopt the Turing model of computation and
we consider the function:
\[
\bez: n,k,A, I_1, \dots, I_k \mapsto \bez (A; I_1, \dots,
I_k)
\]
where all integer numbers are in binary representation,
and $A$ is a list of $n$-tuples 
$(\alpha_1, \dots, \alpha_n)$, and each $I_j$ is a list
of its elements. In particular, the input size is
bounded below by $n \# A_i$ and by $\max_{\boldalpha, i}
\lceil \log_2 \alpha_i \rceil$.
Therefore, $\bez(A; I_1, \cdots, I_k)$ can be computed
in polynomial time by a straight-forward application of
formula (\ref{eq:bezfromI}). As a matter of fact, it
can be computed in time polynomial in the size of $A$.

\begin{problem}[Discrete optimization problem]
\label{prob:opt}
Given $n$ and $A$, compute
\[
\min_{\partition I} \bez(A; {\partition I}) \ ,
\]
where $\partition I =(I_1, \dots, I_k)$ ranges over all the partitions
of $\{1, \dots , n\}$.
\end{problem}

\begin{problem}[Approximation problem]
\label{prob:approx}
Let $C>1$ be fixed. Given $n$ and $A$, compute
some $B$ such that  
\[
BC^{-1} < \min \bez(A; \partition I) < BC
\]
Again, $\partition I = (I_1, \dots, I_k)$ ranges over all the partitions
of $\{1, \dots , n\}$.
\end{problem}

In the problems above, we are not asking for the actual 
partition. 

\begin{restatedtheorem}[restated]
  Problem~\ref{prob:approx} is $\NP$-hard.
\end{restatedtheorem}

This is actually stronger than the conjecture by 
Li and Bai~\cite{LI-BAI}, that corresponds to 
the following immediate
corollary:

\begin{corollary}
  Problem~\ref{prob:opt} is $\NP$-hard.
\end{corollary}

\subsection{Other bounds for the number of roots}
\label{sec:volume}

Kushnirenko's Theorem~\cite{KUSHNIRENKO} bounds
the number of isolated solutions of (\ref{eq:*})
in $(\mathbb C^*)^n$ by $n!\ \vol\ \conv A$,
where $\conv A$ is the smallest convex polytope
containing all the points of $A$. 

This bound is sharper than the B\'ezout bound, but
the known hardness results are far more
dramatic: In~\cite{KHACHIYAN}, Khachiyan proved that
computing the volume of a polytope given by a set
of vertices is $\slashp$-hard. 

There is a large literature on algorithms for approximating
the volume of a convex body given by a separation 
oracle. The problem of approximating the volume of a polytope
in vertex representation can be reduced to the latter by standard
linear programming techniques.

It is known that no deterministic algorithm can approximate the
volume in polynomial time~(\cite{LOVASZ}). However, randomized 
polynomial time
algorithms are known for the same problem~\cites{KLS, WW}.

The same situation seems to be the case regarding the estimation
of the {\em mixed volume}~\cite{DGH}, which gives the actual
number of solutions in $(\mathbb C^ *)^n$ for
generic polynomials of the form (\ref{eq:**})~\cite{BERNSTEIN}.

\subsection{Probabilistic algorithms}
\label{sec:probabilistic}

A {\em probabilistic machine} is a machine that has 
access to {\em random} bits of information, each
random bit costing one unit of time. Each random bit
is an independent, uniformly distributed random variable 
in $\{0,1\}$.  In that sense,
a probabilistic machine is a machine that flips a
fair coin, as many times as necessary, spending
one unit of time at each flip.

We can therefore speak of the probability that the
machine returns a correct result. 

The class $\BPP$
is the class of decision problems $(X, X_{\mathrm{yes}})$ 
such that there is a probabilistic machine and a constant
$\epsilon < 1/2$ that will:
\begin{itemize}
\item[(i)]   Decide in polynomial time if $x \in X$.
\item[(ii)]  Output YES or NO, in polynomial time.
\item[(iii)] For every $x$, the output is the correct
answer to the question: does $x \in X_{\mathrm{yes}}$?
with probability $\ge 1-\epsilon$.
\end{itemize}

Notice that we can improve the probability that the
result is correct by running the same machine several
times. Therefore, in the definition above, we may as
well take $\epsilon = 1/4$.

More generally, a probabilistic machine solves a 
certain problem (e.g. Problem~\ref{prob:approx}) in
polynomial time with probability $\ge 1 - \epsilon$ if
and only if it always terminates in polynomial time,
and the answer is correct with probability $1-\epsilon$.

\begin{restatedtheorem}[restated]
  There is no $\epsilon < 1/2$ and no 
probabilistic machine solving
Problem~\ref{prob:approx} with probability $1-\epsilon$,
unless $\BPP \supseteq \NP$.
\end{restatedtheorem}
\subsection{Approximation classes}
\label{sec:approx}

A theory of complexity classes appropriate for
the study of combinatorial optimization problems
is described in ~\cite{AUSIELLO}. Problem~\ref{prob:opt}
fits naturally in the class of combinatorial 
optimization problems. In this context, 
Problem~\ref{prob:opt} is characterized by:

\begin{itemize}
\item[(i)] A set of {\em instances}, given by the
set of pairs $(n,A)$, 
$n \in \mathbb N, \ A \subset \mathbb N^n$
finite and non-empty. 
\item[(ii)] For every instance $(n,A)$, a set of
feasible solutions, namely the set of partitions
${\partition{I}} = (I_1, \dots, I_k)$ of $\{1, \dots, n \}$. 
\item[(iii)] An objective function (to minimize),
$\bez(A; {\partition I})$. 
\end{itemize}

The class $\NPO$ of combinatorial optimization problems
is analogous to the class $\NP$ of decision problems.
Problem~\ref{prob:opt} belongs to that class:

\begin{itemize}
\item [(1)] The size of each feasible solution is 
polynomially bounded on the size of each instance.
\item [(2)] Given an instance $(n,A)$ and a string $w$,
it can be decided in time polynomial in $(n,A)$ whether
$w$ encodes a feasible solution $\partition I = (I_1, 
\dots, I_k)$.
\item [(3)] The objective function can be computed in
polynomial time.
\end{itemize}

The class $\APX$ of approximable problems in $\NPO$
is defined as the subset of $\NPO$ for which there
is some $C>1$ and a polynomial time algorithm such that,
given an instance of the problem (say $n, A$)) produces
a feasible solution $\partition I$ such that
the objective function applied to that solution
approximates the minimum up to
a factor of $C$.

Theorem~\ref{th:B} admits as a corollary:

\begin{corollary}
  Problem~\ref{prob:opt} does not belong to $\APX$,
unless $\Po = \NP$.
\end{corollary}

Our result actually holds even if we do not require the
algorithms to compute a feasible solution. 

\section{Proof of the Main Theorems}

\subsection{From graph theory to systems of equations.}

\begin{definition}
  A {\em $k$-coloring} of a graph $G=(V,E)$ is
a partition of the set of vertices $V$ into $k$
disjoint subsets (``colors'') $I_j$, so that adjacent
vertices do not belong to the same ``color'' $I_j$.
\end{definition}

\begin{problem}[Graph 3-Coloring] \label{graph3}
Given a graph $G=(V,E)$, decide if there exists a
$3$-coloring of $G$.
\end{problem}

It is known since~\cite{KARP} that the Graph 3-Coloring
Problem is $\NP$-hard (see also~\cite{GAREY-JOHNSON}).
We will actually need to consider an equivalent
formulation of the
Graph 3-coloring problem.

Recall that the cartesian product of two graphs
$G_1 = (V_1, E_1)$ and $G_2 = (V_2, E_2)$ is the
graph $G_1 \times G_2 = (V_1 \times V_2, E)$ with
$( (v_1, v_2), (v_1',v_2') ) \in E$ if and only if
$v_1 = v_1'$ and $(v_2, v_2') \in E_2$ or 
$v_2 = v_2'$ and $(v_1, v_1') \in E_1$.

Also, let $K_3$ denote the complete graph with 3 vertices.

\begin{lemma} \label{lem:equiv}
The graph $G$ admits a 3-coloring if and only if
the graph $G \times K_3$ admits a 3-coloring
$\partition I = (I_1,I_2,I_3)$ with $\# I_1 = \# I_2 = \# I_3 = |G|$.
\end{lemma}

\begin{proof}
$G$ admits a 3-coloring if and only if $G \times K_3$ 
admits a 3-coloring.
Moreover, any coloring $\partition I$
of $G \times K_3$ satisfies 
$\# I_1 = \# I_2 = \# I_3$.
\end{proof}

To each graph $H = (V,E)$ we will associate 
two spaces of polynomial systems. Each of those spaces
is characterized by a support set $A = A(H)$ (resp.
$A(H)^l$) to be constructed and corresponds
to the space of polynomials of the form (\ref{eq:*})
with complex coefficients. Of particular interest will
be graphs of the form $H = G \times K_3$.

We start by identifying the set $V$ of vertices of $H$
to the set $\{1, \dots, m \}$.
Let $K_s$ denote the complete graph of size $s$, i.e.
the graph with $s$ vertices all of them pairwise
connected by edges.

To each copy of $K_s$, $s=0, \dots, 3$ that can
be embedded as a subgraph of $H$ (say the subgraph
generated by $\{v_1, \cdots, v_s\}$) we associate
the monomial
\[
z_{v_1} z_{v_2} \cdots z_{v_s}
\]
\begin{figure}
\centerline{\resizebox{5cm}{!}{\includegraphics{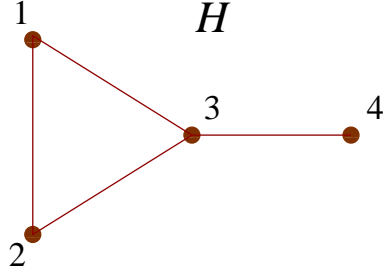}}}
\caption{\label{fig:graph}
In this example, $A(H) = \{ (0,0,0,0), (1,0,0,0), (0,1,0,0),$ 
$(0,0,1,0),$ $(0,0,0,1),$ $(1,1,0,0),$ $(1,0,1,0),$ $(0,1,1,0),$ 
$(0,0,1,1),$ $(1,1,1,0) \}$. A possible polynomial with that support
would be $1+v_1+v_2+v_3+v_4+v_1v_2+v_1v_3 +v_2v_3 + v_3v_4 + v_1v_2v_3$.}
\end{figure}
(the empty graph $K_0$ corresponds to the constant
monomial).
Then we consider the linear space generated by all those
monomials (Figure~\ref{fig:graph}). Therefore, the support 
$A(H)$ is the set
of all 
$e_{v_1} + \cdots + e_{v_s} \subset \mathbb N^m$ such
that $0 \le s \le 3$ and 
$\{v_1, \dots, v_s\}$ induces a copy of $K_s$
as a subgraph of $H$. Here, $e_i$ denotes the $i$-th
vector of the canonical basis of $\mathbb R^n$.

Given a set $A$, we denote by $A^l$ the $l$-fold i
cartesian product of $A$.

The two spaces of polynomial systems associated to
a graph $H$ will be the polynomial systems with
support $A(H)$ and $A(H)^l$.

Remark that none of the two classes of systems 
above is homogeneous in any possible group of
variables (because we introduced a constant
monomial). Therefore, in the calculation of
the B\'ezout number for a partition $\partition I$, 
we can set $a_j = \# I_j$.

\begin{lemma}\label{Aiseasy}
Let $l$ be fixed. Then, there is 
a polynomial time algorithm to compute $A(H)$ and
$A(H)^l$, given $H$.
\end{lemma}

\subsection{A gap between B\'ezout numbers}

In case the graph $H$
admits a $3$-coloring
$\partition I = (I_1, I_2, I_3)$, any corresponding
polynomial system is always trilinear (linear in each
set of variables). If moreover $H$ is of the
form $H=G \times K_3$ with $|G|=n$, the cardinality 
of the
$I_j$ is always $n$, and formula ~(\ref{eq:bezfromI})
becomes: 

\[
\bez (A(G \times K_3); {\partition I}) 
              = \binomial{3n}{n \ n \ n}
\]

The crucial step in the proof of Theorem~\ref{th:A}
is to show that 

\[
\bez (A(G \times K_3); {\partition I}) \ge \frac{4}{3}
\binomial{3n}{n \ n \ n}
\]
unless $k=3$ and ${\partition I}$ is a 3-coloring
of $G \times K_3$.

In order to do that, we introduce the following
cleaner abstraction for the B\'ezout number:
if $k \in \mathbb N$ and ${\mathbf a} 
= (a_1, \dots, a_k) \in \mathbb N^k$
are such that $\sum_{j=1}^k a_j = 3n$, we set
\[
B(\mathbf a) \bydef 
\binomial{3n}{a_1 \ a_2 \ \cdots \ a_k}
\
\prod_{j=1}^k
\left \lceil \frac{a_j}{n} \right \rceil ^{a_j}
\]

\begin{lemma} \label{lem:bezB}
If $H = G \times K_3$ and
${\partition I}=(I_1, \dots, I_k)$ is a partition
of the set $\{1, \dots , 3n\}$ of vertices of $H$, then 
\[
\bez(A(H); {\partition I}) \ge B( \mathbf a)
\]
with $a_j = \# I_j$.
\end{lemma}

\begin{proof}
Consider the $n$ disjoint copies of $K_3$ in
$H=G \times K_3$ induced by the nodes of $G$.
By the pigeonhole principle, 
there is at least one of those copies with
at least $\lceil a_j / n \rceil$ elements of
$I_j$. Hence, the degree $d_j$ in the $j$-th group
of variables is at least $\lceil a_j / n \rceil$.
\end{proof}

The main step towards establishing the ``gap'' is the
following Proposition:

\begin{proposition}\label{prop1}
Let $n, k \in \mathbb N^n$ and let $a_1 \ge a_2 \ge
\cdots \ge a_k \ge 1$ be such that $\sum_{j=1}^k a_j
= 3n$. Then, either $k=3$ and $a_1 = a_2 = a_3 = n$,
or:
\[
B(\mathbf a) \ge \frac{4}{3} B(n,n,n) \ .
\]
Moreover, this bound is sharp.
\end{proposition}

The proof of Proposition~\ref{prop1} is postponed
to section~\ref{sec:prop1}.
\medskip
\par

Putting it all together,

\begin{lemma} \label{lemma:key}
Let $G$ be a graph and $n = |G|$. If $G$ admits a 
3-coloring, then
\[
\min_{\partition I} \bez (A(G \times K_3); \partition I) 
= \binomial{3n}{n \ n \ n}
\]
Otherwise,
\[
\min_{\partition I} \bez (A(G \times K_3); \partition I) \ge \frac{4}{3} 
\binomial{3n}{n \ n \ n}
\]
\end{lemma}

\begin{proof}
According to Lemma~\ref{lem:equiv}, $G$ admits a
3-coloring if and only if $G \times K_3$ 
admits a 3-coloring.

If ${\partition I} = (I_1, I_2, I_3)$ is a 3-coloring of
$G \times K_3$, then
\[
\bez(A(G \times K_3); \partition I)
=
\binomial{3n}{n \ n \ n}
\]

If $\partition I = (I_1, \dots, I_k)$ is not a 3-coloring of
$G \times K_3$, then we distinguish two cases.

We set $a_j = \# I_j$.

{\bf Case 1:} ${\mathbf a}=(n,n,n)$ and hence $k=3$.
Then the degree in at least one group of variables is
$\ge 2$, and
\[
\bez(A(G \times K_3); \partition I)
\ge 2^n
\binomial{3n}{n \ n \ n} 
\]

{\bf Case 2:} ${\mathbf a} \neq (n,n,n)$. Then

\[
\bez(A(G \times K_3); \partition I)
\ge
B(a_1, \dots, a_k)
\ge
\frac{4}{3} 
\binomial{3n}{n \ n \ n}
\ ,
\]
where the first inequality follows from
Lemma~\ref{lem:bezB} and the
second from Proposition~\ref{prop1}.

In both cases,
\[
\min_{\partition I} \bez (A(G \times K_3), \partition I) \ge \frac{4}{3}
\binomial{3n}{n \ n \ n}.
\]
\end{proof}

Lemma~\ref{lemma:key} would be sufficient to prove
a weaker version of Theorem~\ref{th:A}, where the
factor $C$ in problem~\ref{prob:approx} is less 
than $4/3$.

\subsection{Improving the gap}

In order to obtain a proof valid for any $C$
the idea is to increase the gap by considering several copies
of a polynomial system, but each copy in a new set of 
variables. This idea works out because of the special multiplicative
structure of the multi-homogeneous B\'ezout number.
We will need: 

\begin{proposition} \label{prop2}
  Let $m,l \in \mathbb N$. Let $A \subset \mathbb N^m$
be finite and assume that $0 \in A$. Then,
\[
\min_{\partition J} \bez (A^l; \partition J)
=
\binomial{lm}{m \ m \ \cdots \ m} 
\ \left( \min_{\partition I} \bez (A; \partition I) \right)^l
\]
\end{proposition}

\begin{proof}
1. Let $\partition I=(I_1, \cdots, I_k)$ be the partition of 
$\{1, \dots, m\}$ where the minimal B\'ezout
number for $A$ is attained.

This induces a partition $\partition J = (J_{js})_{1\le j\le k, 1 \le s \le l}$ of $\{1, \dots, m\}
\times \{ 1, \dots, l\}$, given by 
$J_{js} = I_j \times \{s\}$. Identifying each pair
$(i,s)$ with $i+ms$, the $J_{js}$ are also a partition
of $\{1, \dots, lm \}$.

By construction of $A^l$, the degree $d_{js}$
in the variables corresponding to $J_{js}$ is
equal to the degree $d_j$ of the variables $I_j$
in $A$. 

The systems corresponding to $A$ and $A^l$ cannot
be homogeneous for any partition, since $0 \in A$ and
$0 \in A^l$. Then we have $a_j = \#I_j = a_{js}$ for any
$s$. Therefore,

\begin{eqnarray*}
\min_{\partition K} \bez (A^l, \partition K)
&\le&
\bez (A^l, \partition J)
\\
&=&
\binomial{lm}
{\underbrace{a_1 \ \cdots \ a_1}_{l \text{\ times}} 
\ \cdots \ 
\underbrace{a_k \ \cdots \ a_k}_{l \text{\ times}}}
\prod_{s=1}^l \prod_{j=1}^k d_j^{a_j} \\
&=&
\binomial{lm}{m\ m\ \cdots \ m}
\
\left( 
\binomial{m}{a_1\ a_2\ \cdots \ a_k}
\prod_{j=1}^k d_j^{a_j} \right)^l \\
&=&
\binomial{lm}{m \ m \ \cdots \ m} 
\ \left( \min_{\partition I} \bez (A; \partition I) \right)^l
\end{eqnarray*}

2. Now, suppose that the minimal B\'ezout number
for $A^l$ is attained for a partition $\partition J=(J_1, 
\cdots, J_r)$.
We claim that each $J_t$ fits into exactly one of
the $l$ sets $\{1, \dots , m\} \times \{s\}$.

Suppose this is not the case. Assume without loss
of generality that $J_1$ splits into $K \subset 
\{1, \dots , m\} \times \{1\}$ and $L \subset
\{1, \dots , m\} \times \{2, \dots, l\}$,
both $K$ and $L$ non-empty.

If $d_K$ denotes the degree in the $K$-variables
and $d_L$ the degree in the $L$ variables, then
$d_1 = d_K + d_L$. Also, $a_1 = a_K + a_L$ where
$a_K$ is the size of $K$ and $a_L$ is the size of $L$.
The multi-homogeneous B\'ezout
number corresponding to the partition ${\partition {J'}} =
(K,L,J_2, \cdots, J_r)$ is:
\[
\bez(A^l; \partition {J' }) =
\binomial{3lm}{a_K \ a_L \ a_2 \ \cdots \ a_r}
d_K^{a_K} d_L^{a_L} \prod_{j=2}^r d_j^{a_j}
\]

Therefore,
\[
\frac
{\bez(A^l; \partition{J' })}
{\bez(A^l, \partition{J})} 
=
\frac {\binomial{a_1}{a_K} d_K^{a_K} d_L^{a_L}}
{(d_K + d_L)^{a_1}}
< 1
\]
and the B\'ezout number was not minimal, thus
establishing the claim.

3. Denote by $\partition J = \cup_{s=1}^l \partition {J^{(s)}}$ 
the partition minimizing
the B\'ezout number corresponding to $A^l$. In the
notation above, we
assume that ${\partition {J^{(s)}}}$ is a partition
of $\{1, \dots, m\} \times \{s\}$.

In that case,
\begin{eqnarray*}
\bez(A^l; \partition J)
&=&
\binomial{lm}{m\ m\ \cdots \ m}
\prod_{s=1}^l
\left(
\binomial{m}{a_{1}^{(s)}\ \cdots \ a_{k}^{(s)}}
\prod_{j=1}^k (d_{j}^{(s)})^{a_{j}^{(s)}}
\right) \\
&=&
\binomial{lm}{m\ m\ \cdots \ m}
\prod_{s=1}^l
\bez(A, \partition{J^{(s)}})
\\
&\ge&
\binomial{lm}{m\ m\ \cdots \ m}
\left( \min_{\partition I} \bez (A ; \partition{I}) \right)^l
\end{eqnarray*}

\end{proof}

Combining Lemma~\ref{lemma:key} and Proposition~\ref{prop2},
we established that:

\begin{lemma} \label{lem:final}
Let $G$ be a graph and $n = |G|$. Let $l \in \mathbb N$.
If $G$ admits a 3-coloring, then
\[
\min_{\partition J} \bez (A(G \times K_3)^l, \partition J) = 
\binomial{3nl}{3n \ 3n \ \cdots \ 3n}
\binomial{3n}{n \ n \ n}^l
\]
Otherwise,
\[
\min_{\partition J} \bez (A(G \times K_3)^l, \partition J) \ge 
\left( \frac{4}{3} \right)^l
\binomial{3nl}{3n \ 3n \ \cdots \ 3n}
\binomial{3n}{n \ n \ n}^l
\]
\end{lemma}

\begin{proof}[Proof of Theorem~\ref{th:A}]
Assume that {\tt ApproxB\'ez} is a deterministic,
polynomial time algorithm for solving
problem~\ref{prob:approx}, i.e., for
estimating the B\'ezout number up to
a factor of $C$.

Then the following algorithm decides
Graph 3-coloring (Problem~\ref{graph3})
in polynomial time:

\begin{algorithm}[Decides Graph 3-coloring problem]
\label{alg1}
\end{algorithm}
\begin{trivlist}
\item [] {\tt Input: a graph} $G$ {\tt of size} $n$.
\item [] {\tt Output: } YES if $G$ admits a 3-coloring, NO otherwise.
\item [] {\tt Constants: } 
$l = 
\left \lceil \frac{\log C }{2 \log 4/3} \right \rceil$.
\item [\text{1.}] {\tt Compute } 
\[
\rho \leftarrow
\frac{ {\text{\sc ApproxB\'ez}}
\left(A(G \times K_3)^l\right) }
{
\binomial{3nl}{3n \ 3n \ \cdots \ 3n}
\binomial{3n}{n \ n \ n}^l
}
\]
\item [\text{2.}] {\tt \bf If }$\rho^2 < C$ 
{\tt {\bf then} Output YES, 
{\bf else} Output NO.}
\end{trivlist}

By our choice of the constant $l$, $\sqrt{C} \le (4/3)^l$.
Therefore, Lemma~\ref{lem:final} asserts that the output
of algorithm~\ref{alg1} is correct.

The bit-size of the numbers that occur when computing
the denominator of line 2 are bounded above by
$O(3nl \log (3nl))$. The size of the graph $G \times K_3$
is $O(n)$, and Lemma~\ref{Aiseasy} says that $A^l$ can
be computed in polynomial time.

It follows that Algorithm~\ref{alg1} runs in polynomial
time. Since Graph 3-coloring is $\NP$-complete, we
deduce that $\Po = \NP$.
\end{proof}

\begin{proof}[Proof of Theorem~\ref{th:B}]
Assume now that {\tt ApproxB\'ez} is a probabilistic 
polynomial time algorithm for solving
problem~\ref{prob:approx}, which returns a correct
result with probability $1-\epsilon$, $\epsilon < 1/4$.

Then Algorithm~\ref{alg1} will return the correct answer
for the Graph 3-coloring Problem, with probability at
least $1-\epsilon$. This implies that 
Problem~\ref{graph3} is actually in $\BPP$. 
\end{proof}

\section{Proof of Proposition \ref{prop1}}
\label{sec:prop1}

We will need the following trivial Lemma in the proof of
Proposition~\ref{prop1}:

\begin{lemma}\label{lem:triv}
  Let $x, n \in \mathbb N$. Then,
\[
\left( \left\lceil \frac{x}{n} \right\rceil \frac{n}{x}
\right)^x
\ge
1 + \left(\rule{0em}{2ex} (n-x) \mod n \right)
\ .\]
In particular, the left-hand side is $\ge 2$ whenever
$n \not \: \mid x$, and is always $\ge 1$.
\end{lemma}

\begin{proof} Since $n \left\lceil \frac{x}{n} \right\rceil
= x + (n-x) \mod n$,
we have:
\[
\left( \left\lceil \frac{x}{n} \right\rceil \frac{n}{x}
\right)^x
=
\left( 
1 + \frac{(n-x) \mod n }{x}
\right)^x
\ge
1 + \left(\rule{0em}{2ex} (n-x) \mod n \right)
\]
\end{proof}

  Also, we will make use of the Stirling 
Formula~\cite[(6.1.38)]{HANDBOOK}:
\begin{equation} \label{Stirling}
x! = \sqrt{2 \pi} \ x^{x+\frac{1}{2}} \ 
e^{-x+ \frac{\theta(x)}{12x}} \ ,
\end{equation}
where $0 < \theta(x) < 1$.

\begin{proof}[Proof of Proposition~\ref{prop1}]
The ratio between $B(\mathbf a)$ and
$B(n,n,n)$ is:
\[
\frac{B(\mathbf a)}{B(n,n,n)}
=
\prod_{j=1}^k
\left\lceil \frac{a_j}{n} \right\rceil^{a_j}
\
\frac{ n! \ n! \ n! } {a_1! \ a_2! \ \cdots \ a_k!}
\]

From Stirling formula~(\ref{Stirling}) it follows 
immediately that:

\begin{equation}\label{ratio}
\frac{B(\mathbf a)}{B(n,n,n)}
=
\sqrt{2\pi}^{3-k}
\
\prod_{j=1}^k 
\left\lceil \frac{a_j}{n} \right\rceil^{a_j}
\
\frac{ n^{3n + \frac{3}{2}} }
{ \prod_{j=1}^k a_j^{a_j + \frac{1}{2}} }
\
e^{ \frac{\theta(n)}{4n}-\sum \frac{\theta(a_j)}{12 a_j}}
\end{equation}

Now we distinguish the cases $k=1$, $k=2$, and $k \ge 3$.
The first two cases are easy:
\medskip

{\bf Case 1:} If $k=1$, then $a_1 = 3n$ and (\ref{ratio})
becomes:
\[
\frac{B(\mathbf a)}{B(n,n,n)}
=
2 \pi \ \frac{n}{\sqrt{3}} \
e^{ \frac{\theta(n)}{4n}-\frac{\theta(3n)}{36n}}
\]
which is bounded below by $\frac{2 \pi}{\sqrt{3}} e^{-1/36}
\simeq 3.528218766$.
\medskip

{\bf Case 2:} If $k=2$, Lemma~\ref{lem:triv} implies
that
\[
\frac{B(\mathbf a)}{B(n,n,n)}
\ge
\sqrt{2 \pi} \ \frac{n^{\frac{3}{2}}}{\sqrt{a_1 \ a_2}}
\ e^{-1/6}
\]

Since $\sqrt{a_1 \ a_2} \le \frac{a_1 + a_2}{2} 
= \frac{3n}{2}$, we obtain:

\[
\frac{B(\mathbf a)}{B(n,n,n)}
\ge
\frac{2}{3} \sqrt{2 \pi} e^{-1/6} \simeq 
1.414543350
\]
\medskip

{\bf Case 3:} Let $k \ge 3$. If $a_3 = n$, then
$k=3$ and $a_1 = a_2 = a_3 = n$, so there is nothing
to prove. Therefore, we assume from now on that $a_3 < n$.

We separate the right-hand side of (\ref{ratio})
into two products, the first for $j=1,2,3$ and the
second for $j \ge 4$. Equation (\ref{ratio}) becomes
now:

\begin{equation}\label{ratio2}
\begin{split}
\frac{B(\mathbf a)}{B(n,n,n)}
=
\left(
\prod_{j=1}^3 
\left(
\left\lceil \frac{a_j}{n} \right\rceil
\frac{n}{a_j}
\right)^{a_j}
\
\frac{ n^{\frac{3}{2}} }
{ \sqrt{a_1 a_2 a_3 }}
\
e^{ \frac{\theta(n)}{4n}-\sum_{j=1}^{3}
\frac{\theta(a_j)}{12 a_j}}
\right)
\\
\left(
\sqrt{2\pi}^{3-k}
\
\prod_{j=4}^k 
\frac{n^{a_j}}{a_j^{a_j + \frac{1}{2}}}
\
e^{ -\sum_{j=4}^k \frac{\theta(a_j)}{12 a_j}}
\right)
\end{split}
\end{equation}
using the fact that $a_j < n$ for $j \ge 4$. In case
$k=3$, the second factor in equation (\ref{ratio2}) above is 
equal to one.

Since $a_3 < n$, $n \not \: \mid a_3$ and
Lemma~\ref{lem:triv} implies that for $a_3 < n$
\[
\prod_{j=1}^3 
\left(
\left\lceil \frac{a_j}{n} \right\rceil
\frac{n}{a_j}
\right)^{a_j}
\ge 2
\]

Moreover, $\sqrt[3]{a_1 a_2 a_3} \le (a_1+a_2+a_3)/3 \le n$,
so the first factor of the right-hand side of
(\ref{ratio2}) can be bounded below by
\[
\prod_{j=1}^k 
\left(
\left\lceil \frac{a_j}{n} \right\rceil
\frac{n}{a_j}
\right)^{a_j}
\
\frac{ n^{\frac{3}{2}} }
{ \sqrt{a_1 a_2 a_3 }}
\
e^{ \frac{\theta(n)}{4n}-\sum_{j=1}^{3}
\frac{\theta(a_j)}{12 a_j}}
\ge
2 e^{-1/4}
\simeq
1.557601566
\]

If $k=3$ we are done. Otherwise, we notice that
since the $a_j$ are non-increasing, 
$a_j \le \frac{3n}{4}$ for all $j \ge 4$. In order
to bound the second factor of (\ref{ratio2}),
we will need the following technical Lemma:

\begin{lemma} \label{case-a-small}
Let $n,x \in \mathbb N$ and let 
$x \le \frac{3n}{4}$. Then,
unless $(n,x) \in \{(2,1),(3,2),(4,3),\\
(6,4),(7,5),(8,6)\}$, we have:
\[
\frac{ n^{x} } {\sqrt{2 \pi} x^{x + \frac{1}{2}}} 
e^{-\frac{1}{12x}}
>
1
\]
\end{lemma}

(Proof is postponed).

Therefore, unless some of the pairs $(n, a_j)$,
$j \ge 4$ belong to the exceptional subset
$\{(2,1),(3,2),(4,3), (6,4),(7,5),(8,6) \}$, we
have:
\[
\frac{B(\mathbf a)}{B(n,n,n)}
\ge
2 e^{-\frac{1}{4}} \simeq 1.557601566 \ .
\]

Finally, we consider the values of $n$ and $\mathbf a$
where some $(n,a_j)$, $j \ge 4$,
is in the exceptional subset. All the possible values
of $n$ and $\mathbf a$ are listed in table~\ref{tab1}.
\begin{table}
\centerline{
\begin{tabular}{||c|c|c|l|c|r|r|c||}
\hline
\hline
$n$ & $a_j$& $3n$ & $\mathbf a$ & j 
& $B(\mathbf a)$ & $B(n,n,n)$& 
$\frac{B(\mathbf a)}{B(n,n,n)}$\\
\hline
\hline
2 & 1 & 6 & 1 1 1 {\bf 1} {\bf 1} {\bf 1}
                              & 4,5,6 
			          &          720 & 90
				  & 8\\
  &   &   & 2 1 1 {\bf 1} {\bf 1} 
                              & 4,5 
			          &          360 &
				  &4\\
  &   &   & 2 2 1 {\bf 1}     & 4 &          180 &
                                  &2\\
  &   &   & 3 1 1 {\bf 1}     & 4 &          120 &
                                  &$\frac{4}{3}$\\
\hline
3 & 2 & 9 & 2 2 2 {\bf 2} 1   & 4 &        22680 & 1680
                                  &$\frac{27}{2}$\\
  &   &   & 3 2 2 {\bf 2}     & 4 &         7560 &
                                  &$\frac{9}{2}$\\
\hline
4 & 3 &12 & 3 3 3 {\bf 3}     & 4 &       369600 & 34650
                                  &$\frac{32}{3}$\\
\hline
6 & 4 & 18& 4 4 4 {\bf 4} 1 1 & 4 &  19297278000 & 17153136
                                  &1125\\
  &   &   & 4 4 4 {\bf 4} 2   & 4 &   9648639000 &
                                  &$\frac{1125}{2}$\\
  &   &   & 5 4 4 {\bf 4} 1   & 4 &   3859455600 &
                                  &225\\
  &   &   & 5 5 4 {\bf 4}     & 4 &    771891120 &
                                  &45\\
  &   &   & 6 4 4 {\bf 4}     & 4 &    643242600 &
                                  &$\frac{75}{2}$\\
\hline
7 & 5 & 21& 5 5 5 {\bf 5} 1   & 4 & 246387645504 & 399072960
                                  &$\frac{3087}{5}$\\
  &   &   & 6 5 5 {\bf 5}     & 4 &  41064607584 &
                                  &$\frac{1029}{10}$\\
\hline
8 & 6 & 24& 6 6 6 {\bf 6}  & 4 & 2308743493056 & 9465511770 
&$\frac{10976}{45}$\\
\hline
\hline
\end{tabular}
}
\caption{\label{tab1}
Ratios for all the exceptional pairs $(n, \mathbf a)$.}
\end{table}
The ratio is always $\ge 4/3$, and the value of $4/3$
is attained for $n=2$ and $\mathbf a = (3,1,1,1)$.
\end{proof}

\begin{proof}[Proof of Lemma~\ref{case-a-small}]
\begin{figure}
\includegraphics{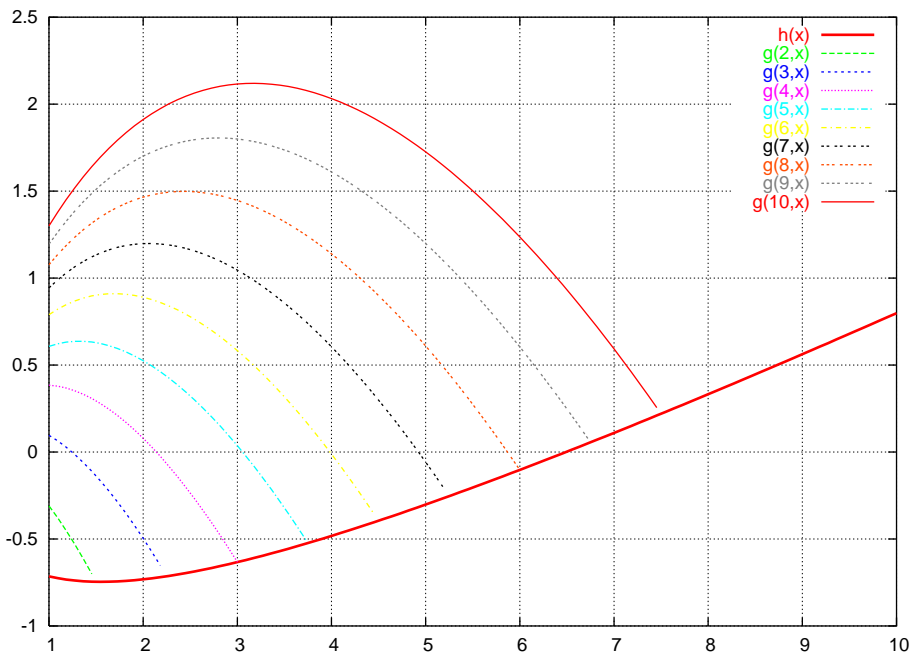}
\caption{\label{fig-gandh}Plots of $g_n(x)$ and $h(x)$.}
\end{figure}

Let 
\begin{eqnarray*}
g_n(x) &=& \log \left(
\frac{ n^x }{\sqrt{2 \pi} x^{x + \frac{1}{2}}} 
e^{-\frac{1}{12x}} \right) \\
&=&
x \log n 
- x \log x 
- \frac{1}{2} \log x
- \frac{1}{12x} 
- \frac{1}{2} \log 2\pi
\end{eqnarray*}

(see figure~\ref{fig-gandh}).
We first consider values of $x \ge 7$.
By hypothesis, $n/x \ge 4/3$ so 
$\log n - \log x \ge \log(4/3)$, and therefore $g_n(x) 
\ge h(x)$, where:
\[
h(x) = x \log(4/3) 
- \frac{1}{2} \log x
- \frac{1}{12x}
- \frac{1}{2} \log 2\pi
\]

(see Figure~\ref{fig-gandh} also). 
Notice that $h(x)$ is independent of $n$.
The derivative of $h$ is
\[
h '(x) = 
\log(4/3) 
- \frac{1}{2x} 
+ \frac{1}{12x^2}
=
\frac{12 \log(4/3) x^2 - 6x + 1}{12x^2}
\]

The numerator vanishes at
\[
x = \frac{1 \pm 1\sqrt{1-4/3 \log(4/3)}}{4 \log(4/3)}
\]
Numerically, $x \simeq 0.1867281114$ or 
$x \simeq 1.551301638$. 
Therefore, the function $h(x)$ is increasing for $x \ge 2$.
Again, numerically $h(7) \simeq 0.1099761345$ and therefore,
if $x \ge 7$ we always have:
\[
e^{g_n(x)} \ge 1.1162 > 1 
\]

Now we consider $x \le 6$. 
Having $g_n > 0$,
is equivalent to:
\[
n  > n_0(x) =
x 
e^{\frac{1}{2x} \log x
+\frac{1}{12x^2} 
+\frac{1}{2x} \log 2\pi}
\]

At this point, we proved that 
$g_n(x)$ is positive, except possibly for
pairs $(n,x)$ with $1 \le x \le 6$ and 
$\frac{4}{3}x \le n \le n_0(x)$. The values
of $n_0$ are tabulated in Table~\ref{tabn0}.
From Table~\ref{tabn0} it is clear that the
only exceptions are those listed
in the hypothesis.

\begin{table}[ht]
\centerline{
\begin{tabular}{||l|l|l|c||}
\hline
\hline
$x$ & $\frac{4}{3}x$& $n_0(x)$    &Possible $n$'s \\
\hline
1   & 1.333333333  & 2.724464424 &  2  \\
2   & 2.666666666  & 3.844857634 & 3 \\
3   & 4            & 4.939610298 & 4 \\
4   & 5.333333333  & 6.016610872 & 5 \\
5   & 6.666666666  & 7.081620345 & 6 \\
6   & 8            & 8.137996302 & 8 \\
\hline
\hline
\end{tabular}}
\caption{\label{tabn0}Possible values of $n$ for $x$ small}
\end{table}

\end{proof}


\end{document}